\documentclass[sigconf]{acmart}

\copyrightyear{2025}
\acmYear{2025}
\acmConference[UbiComp Companion '25]{Companion of the 2025 ACM International Joint Conference on Pervasive and Ubiquitous Computing}{October 12--16, 2025}{Espoo, Finland}
\acmBooktitle{Companion of the 2025 ACM International Joint Conference on Pervasive and Ubiquitous Computing (UbiComp Companion '25), October 12--16, 2025, Espoo, Finland}\acmDOI{10.1145/3714394.3756345}
\acmISBN{979-8-4007-1477-1/2025/10}

\usepackage{tcolorbox}
\usepackage{makecell}
\usepackage{epstopdf} 
\usepackage{soul}
\usepackage{graphicx}
\usepackage{subfigure}
\usepackage{cleveref}
\usepackage{adjustbox}
\usepackage{multirow}
\usepackage{ulem}

\definecolor{mygreen}{RGB}{0,128,0}
\definecolor{myred}{RGB}{255,0,0}
\begin{document}

\title{Behavioral Indicators of Loneliness: Predicting University Students' Loneliness Scores from Smartphone Sensing Data}

\author{Qianjie Wu}
\orcid{0009-0008-7175-0171}
\affiliation{%
  \institution{The Hong Kong Polytechnic University}
  \city{Hong Kong}
  \country{China}}
\email{qian-jie.wu@connect.polyu.hk}

\author{Tianyi Zhang}
\orcid{0000-0002-0778-8844}
\affiliation{%
  \institution{The University of Melbourne}
  \city{Melbourne}
  \country{Australia}}
\email{t.zhang59@student.unimelb.edu.au}

\author{Hong Jia}
\orcid{0000-0002-6047-4158}
\affiliation{%
  \institution{University of Auckland}
  \city{Auckland}
  \country{New Zealand}}
\email{hong.jia@auckland.ac.nz}

\author{Simon D'Alfonso}
\orcid{0000-0001-7407-8730}
\affiliation{%
  \institution{The University of Melbourne}
  \city{Melbourne}
  \country{Australia}}
\email{dalfonso@unimelb.edu.au}

\renewcommand{\shortauthors}{Qianjie Wu, Tianyi Zhang, Hong Jia \& Simon D’Alfonso}
\renewcommand{\shorttitle}{Smartphone Behavioral Indicators of University Students' Loneliness}

\begin{abstract}
Loneliness is a critical mental health issue among university students, yet traditional monitoring methods rely primarily on retrospective self-reports and often lack real-time behavioral context. This study explores the use of passive smartphone sensing data to predict loneliness levels, addressing the limitations of existing approaches in capturing its dynamic nature. We integrate smartphone sensing with machine learning and large language models respectively to develop generalized and personalized models. Our Random Forest generalized models achieved mean absolute errors of 3.29 at midterm and 3.98 (out of 32) at the end of semester on the UCLA Loneliness Scale (short form), identifying smartphone screen usage and location mobility to be key predictors. The one-shot approach leveraging large language models reduced prediction errors by up to 42\% compared to zero-shot inference. The one-shot results from personalized models highlighted screen usage, application usage, battery, and location transitions as salient behavioral indicators. These findings demonstrate the potential of smartphone sensing data for scalable and interpretable loneliness detection in digital mental health.

\end{abstract}




\maketitle

\section{Introduction}
With the increasing prevalence of loneliness among contemporary youth populations, university students have emerged as a high-risk group for experiencing loneliness due to academic pressures, evolving social dynamics, and increasing individualism \citep{barreto2021loneliness}. Continuous monitoring of loneliness has become a critical prerequisite for identifying potential psychological crises and implementing targeted interventions. However, existing assessment methods often fail to capture the dynamic and fluctuating nature of loneliness in real-world settings. For example, self-reported assessments require frequent manual administration, which can create substantial data cleaning challenges for researchers and contribute to participant fatigue \citep{hays1987short}.

In this context, digital phenotyping offers a promising approach to dynamically modeling loneliness by continuously capturing device usage and behavioral patterns \citep{onnela2016harnessing}. With their ubiquity and ability to passively record rich behavioral and social interaction data, smartphones enable objective tracking of daily activities, supporting scalable behavioral science research \citep{harari2017smartphone}.

Prior studies have explored the predictive power of smartphone sensing features for estimating loneliness scores. One study \citep{doryab2019identifying} developed a machine learning model to classify loneliness using smartphone and Fitbit data, achieving an accuracy of 80.2\% for binary classification tasks. Similarly, another study \citep{qirtas2023personalizing} improved model performance through behavioral grouping (F1 = 92.5\%). While prior studies have explored loneliness prediction using deep neural networks or ensemble learning, these approaches face challenges of personalization due to limited data availability. As a result, there is a critical need for advanced approaches that can operate effectively on small datasets, enhance contextual understanding, and identify the most predictive smartphone sensors and features, ultimately improving model interpretability and alignment with real-world behavioral dynamics.

To address the gap, our study employs advanced approaches to predict short-form UCLA Loneliness Scale (ULS-8) scores for 88 university students using multi-dimensional smartphone sensing data. To explore the best smartphone sensors and features for such tasks, we established generalized predictive models using Random Forest, which predicts ULS-8 loneliness scores by extracting multiple features from smartphone usage data. These models achieved mean absolute error of 3.29 and 3.98 (out of 32) at midterm and the end of semester respectively, with features transitions between screen unlock duration, distinct locations, and screen unlock episode count emerging as strong predictors. To explore the potential for interpretable and contextual-aware analysis of smartphone-sensed behavioral information in predicting loneliness levels, and to address the challenge of limited or missing data, we also constructed personalized predictive models leveraging large language models. The one-shot approach highlighted screen, applications, battery, and locations as more salient behavioral indicator sensors, demonstrating the potential to interpret context-specific signals in a personalized manner.

\section{Method and Experiments}
\subsubsection{\textbf{Participants and Data Collection}}
To investigate the ability to predict loneliness in university students using passive data collected from smartphones, data collection was implemented via the AWARE application \citep{van2023aware}. In this study, we collected smartphone sensor data and the ULS-8 loneliness scores over a university semester (17 weeks). The following sensor data were collected: Application Usage, Battery, Calls, Keyboard, Locations, Messages, and Screen. The ULS-8 scale was delivered at two time points: midterm (Week 7) and end-of-semester (Week 17). The questionnaire includes 8 items, with 6 positively scored and 2 reverse scored items \citep{hays1987short}. Responses use a 4-point scale (1 = ``never'' and 4 = ``always''), yielding total scores ranging from 8 to 32, with higher scores indicating greater loneliness. The individual items and scale details are described in Figure~\ref{fig:zero-shot}. The study was approved by the University of Melbourne’s Office of Research Ethics and Integrity (approval number: 2024-26643-60526-9).

\subsubsection{\textbf{Data Processing}}

For the ULS-8 scores, we computed the total score according to the questionnaire instructions, and treated each individual item score separately. Meanwhile, we used smartphone sensing features from the two weeks prior to each ULS-8 assessment to capture recent behavioral patterns that are most indicative of measured levels of loneliness. The selection of a preceding two-week period is supported by prior studies linking short-term behavioral data with mental health outcomes \citep{yang2023loneliness, wu2021improving}. Feature extraction at daily granularity was performed using the RAPID tool \citep{rapids}, generating 76 raw features across smartphone sensor dimensions. In data cleaning, we retained the participant data only if their duration covered at least seven days within the 14-day observation period. Of the 151 participants in the study, the data from 88 students were deemed valid for analysis after excluding dropouts and applying screening criteria. For machine learning modeling, the 76 features were mapped across 14 days, yielding a total of 1064 daily-granularity features.

Additionally, we generated textual descriptions for these smartphone sensing features derived from 7 different sensors for use in the large language model (LLM) analysis, including the following:

\begin{itemize}
    \item Applications: Features related to the usage of various phone applications, such as the total duration and number of usage episodes for all applications, social media, dating, entertainment, and other specific apps like Facebook Moments, YouTube, and Twitter.

    \item Battery: Features capturing the number and duration of battery charging and discharging episodes.

    \item Calls: Features tracking phone call activities, including the number of incoming, outgoing, and missed calls, as well as their duration and the contacts involved.

    \item Keyboard: Features analyzing typing behaviors, such as the number of key presses, changes in text length, average session length, and the average time between keystrokes.

    \item Locations: Features related to the phone's location data, such as time spent at different locations, travel distance, speed, and various metrics indicating the user's movement and location entropy.

    \item Messages: Features associated with the sending and receiving of messages, including the number of messages, distinct contacts involved, and interaction with the most frequent contacts.

    \item Screen: Features tracking the phone's screen unlock behavior, including the number of unlock episodes, duration of unlock sessions, and the times between unlocks.
\end{itemize}

Each smartphone feature description in LLM inferences was formatted as: ``\textit{<feature description>, <day 1 value>, <day 2 value>, ..., <day 7 value>} (weekly average \textit{<average value>})".

\subsubsection{\textbf{Generalized Models}}

To reveal dynamic associations among features from different sensors and the reported loneliness scale, random forest regression models were used within a Recursive Feature Elimination (RFE) framework to quantify the contribution of sensor features to ULS-8 item prediction. Prediction performance of Mean Absolute Error of feature subsets was dynamically evaluated via 3-fold cross-validation. In each iteration, the least important feature (ranked by Gini index) was removed, and the decay trajectory of model performance during dimensionality reduction was recorded, generating a feature contribution ranking map and dynamic elimination log.

\subsubsection{\textbf{Task Description for Personalized Models}}

Leveraging LLMs to understand the association between multi-sensor behavioral features of smartphones and loneliness phenotypes, this study designed zero-shot and one-shot prediction experiments for the Week 17 end-of-semester stages for each sensor. Our prompts required the model to output the predicted scores (ranging from 1 to 4) of each individual ULS-8 items and a textual reasoning for the predicted score.

We evaluated the inference performance by testing each sensor individually, introducing one sensor at a time in both zero-shot and one-shot inferences. These more focused inferences were to better isolate the contribution of each sensor modality since our goal was to identify in which sensors are the most effective for such tasks. In the zero-shot inference approach, we provided daily structured natural language descriptions of smartphone trace features over a two-week observation period and defined tasks for the LLMs to infer predictive scores of the ULS-8 scale. In the one-shot inference approach, we provided the LLMs with a comprehensive input for each prediction instance, including smartphone sensor data from the two weeks preceding the midterm checkpoint and the corresponding ULS-8 item scores. It was then given behavioral data from the two weeks before the end-of-semester stage and tasked with predicting the ULS-8 scores based on that data.

To minimize output randomness, the model's temperature parameter was set to zero. The experiment was conducted using the Gemini 2.0 flash model. No other LLMs were included since comparative evaluation of multiple LLMs was outside the scope of this work. The zero-shot and one-shot prompt templates are shown in Figure \ref{fig:zero-shot} and Figure \ref{fig:one-shot}.

\begin{figure}[htbp]
\centering
\fbox{%
    \parbox{1\columnwidth}{
    \fontsize{6}{7}\selectfont
    You are an expert in analysing human behavior and psychological wellbeing. The following is a summary of a university student's smartphone usage data, collected via \textit{\{sensor\}}. These activity metrics were recorded daily over the weeks preceding the administration of the UCLA Loneliness Scale questionnaire. \\

    Please review the data and, based on the behavioral patterns reflected in \textit{\{sensor\}}, infer the student's responses to each item on the UCLA Loneliness Scale (short form). Use a scale from 1 to 4, where: \\
    \hspace*{1em}• 1 = Never \\
    \hspace*{1em}• 2 = Rarely \\
    \hspace*{1em}• 3 = Sometimes \\
    \hspace*{1em}• 4 = Often \\

    Your reasoning for each score should be concise (no more than two sentences) and grounded in the behavioral data provided. \\

    \textbf{Smartphone Usage Data (Daily Records):} \\
    \textit{\{feature description\}} \\

    \vspace{0.5em}
    \textbf{Task:} Based on the data above, complete the UCLA Loneliness entries (items 1--8) using scales 1 to 4 and a brief rationale grounded in the behavioral features: \\

    \textbf{Items:} \\
    1. I lack companionship. \\
    2. There is no one I can turn to. \\
    3. I am an outgoing person. \\
    4. I feel left out. \\
    5. I feel isolated from others. \\
    6. I can find companionship when I want it. \\
    7. I am unhappy being so withdrawn. \\
    8. People are around me but not with me. \\

    \vspace{0.5em}
    \textbf{Response Format:} \\
    \texttt{[} \\
    \texttt{\{"entry": 1, "score": <1--4>, "reason": "<2 sentences max>"\},} \\
    \texttt{\{"entry": 2, "score": <1--4>, "reason": "<2 sentences max>"\},} \\
    \texttt{...} \\
    \texttt{\{"entry": 8, "score": <1--4>, "reason": "<2 sentences max>"\}} \\
    \texttt{]}
    }
}
\caption{Zero-shot prompt instruction for LLMs to predict the ULS-8 loneliness score using smartphone usage data.}
\label{fig:zero-shot}
\end{figure}

\begin{figure}[htbp]
\centering
\fbox{%
    \parbox{0.97\columnwidth}{
    \fontsize{6}{7}\selectfont
    You are an expert in analysing human behavior and psychological wellbeing. Below is a university student’s smartphone usage summary collected via \textit{\{sensor\}} over two weeks prior to completing the UCLA Loneliness Scale (ULS-8, short form). Each week’s data includes daily activity metrics. \\

    Your task is to review this behavioral data and predict the student’s likely responses to each of the 8 ULS-8 items. Use the 4-point scale: \\
    \hspace*{1em}• 1 = Never \\
    \hspace*{1em}• 2 = Rarely \\
    \hspace*{1em}• 3 = Sometimes \\
    \hspace*{1em}• 4 = Often \\

    Provide a numerical score for each item along with a concise rationale (maximum 2 sentences) based on the behavioral trends in the data. \\

    \textbf{ULS-8 Items:} \\
    1. I lack companionship. \\
    2. There is no one I can turn to. \\
    3. I am an outgoing person. \\
    4. I feel left out. \\
    5. I feel isolated from others. \\
    6. I can find companionship when I want it. \\
    7. I am unhappy being so withdrawn. \\
    8. People are around me but not with me. \\

    \vspace{0.3em}
    \textbf{Example:} \\
    The following data presents earlier behavioral data from the same individual, collected approximately half a university semester before the current data. It demonstrates how similar behavioral patterns corresponded to their ULS-8 responses at that time. \\

    Daily activity metrics from Day 1 to Day 7 of the first week preceding the administration of the UCLA Loneliness Scale questionnaire: \\
    \textit{\{feature description\}} \\

    Daily activity metrics from Day 1 to Day 7 of the second week preceding the administration of the UCLA Loneliness Scale questionnaire: \\
    \textit{\{feature description\}} \\

    The corresponding ULS-8 item scores were: \textit{<Q1 score>, <Q2 score>, ..., <Q8 score>} (for entries 1 to 8) \\

    \vspace{0.5em}
    \textbf{Task:} Based on the earlier example, please now assess the current smartphone activity data from the same student, recorded in the week immediately preceding their latest UCLA Loneliness Scale questionnaire. Note: Some zero entries may correspond to missing values and should be interpreted with caution. \\

    Daily activity metrics from Day 1 to Day 7 of the first week preceding the administration of the UCLA Loneliness Scale questionnaire: \\ 
    \textit{\{feature description\}} \\

    Daily activity metrics from Day 1 to Day 7 of the second week preceding the administration of the UCLA Loneliness Scale questionnaire: \\ 
    \textit{\{feature description\}} \\

    \vspace{0.5em}
    \textbf{Response Format:} \\
    \texttt{[} \\
    \texttt{\{"entry": 1, "score": <1--4>, "reason": "<2 sentences max>"\},} \\
    \texttt{\{"entry": 2, "score": <1--4>, "reason": "<2 sentences max>"\},} \\
    \texttt{...} \\
    \texttt{\{"entry": 8, "score": <1--4>, "reason": "<2 sentences max>"\}} \\
    \texttt{]}
    }
}
\caption{One-shot prompt instruction for LLMs to predict the ULS-8 loneliness score using smartphone sensor data.}
\label{fig:one-shot}
\end{figure}

\subsubsection{\textbf{Evaluation}}

We evaluated model performance using Mean Absolute Error (MAE) measured average prediction error and Mean Bias Error (MBE) captured directional bias. These metrics were applied across both machine learning generalized models and LLM-based personalized inference models.

For evaluating the total ULS-8 score in personalized models, we summed the predicted scores of individual ULS-8 items for each participant and performed the evaluation across all participants.

\section{Results \& Discussion}
\subsection{Personalized LLMs Inferences for Predicting Loneliness Scores}
The zero-shot MAE results shown in Table \ref{tab:sensor_performance} indicates limited predictive performance of the ULS-8 scale when using information provided by individual sensors (MAE $\geq 7.4$), with performance varying across different sensor types. Specifically, the battery and screen sensors demonstrate the best predictive performance in the zero-shot scenario (MAE = 7.4), while the keyboard sensor shows the lowest performance (MAE = 11.1). These differences may be attributed to the varying degrees to which LLMs process subjective factors expressed by different sensor data, leading to different dependencies on prior knowledge.

\begin{figure}[ht]
    % \centering
    {
        \includegraphics[width=\linewidth]{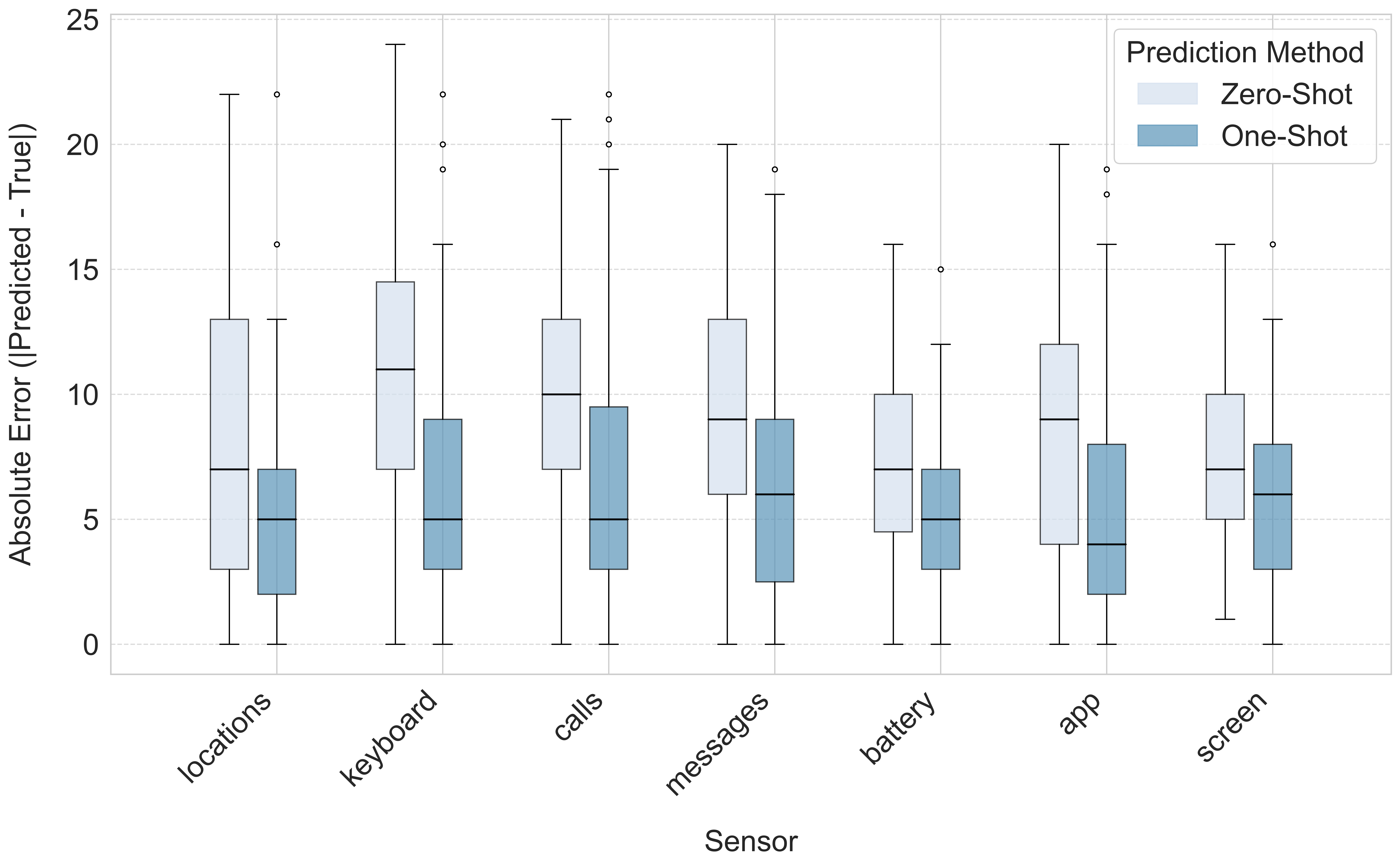}
    }
    \vspace{-5mm}
    \caption{Comparison of feature extraction errors across smartphone sensors using LLMs, averaged by participants’ total ULS-8 loneliness scores (N = 88).}
    \label{box_plot}
\end{figure}

As shown in Figure \ref{box_plot}, the one-shot prediction results consistently outperforms those of zero-shot prediction across all sensors. The interquartile range (IQR) is significantly reduced in the one-shot mode (e.g., the IQR of app sensors narrows from [4-12] to [2-8]), indicating improved prediction stability. Notably, the keyboard sensor showed the greatest improvement, with a decrease in median absolute error from 11 in the zero-shot setting to 5 with one-shot learning, and a decrease of 42\% MAE from 11.1 to 6.4 (Table \ref{tab:sensor_performance}).

\begin{table}[htbp]
\centering
\small
\caption{LLM Predictive Performance Comparison: ULS-8 Total Metrics for Each Sensor in Zero-shot vs. One-shot Mode}
\label{tab:sensor_performance}
\scalebox{0.85}{
\begin{tabular}{lcc|cc|cc}
\toprule
\textbf{Sensor} & \multicolumn{2}{c}{\textbf{Zero-shot}} & \multicolumn{2}{c}{\textbf{One-shot}} & \multicolumn{2}{c}{\textbf{Change Rate}}\\
\cmidrule(lr){2-3} \cmidrule(lr){4-5} \cmidrule(lr){6-7}
& MAE & MBE & MAE & MBE & MAE & MBE \\
\hline
Applications & 8.60 & 8.32 & 5.17 & 3.87 & -39.9\% & -53.4\%\\
Battery      & 7.44 & 6.84 & 5.27 & 3.65 & -29.2\% & -46.6\%\\
Calls        & 9.67 & 8.97 & 7.21 & 6.16 & -25.5\% & -31.3\%\\
Keyboard     & 11.10 & 10.90 & 6.40 & 5.10 & -42.3\% & -53.3\%\\
Locations    & 8.11 & 7.13 & 5.59 & 4.16 & -31.1\% & -41.6\%\\
Messages     & 9.33 & 8.86 & 6.24 & 5.06 & -33.2\% & -42.8\%\\
Screen       & 7.44 & 6.71 & 5.35 & 3.13 & -28.1\% & -53.4\%\\
\bottomrule
\end{tabular}
}
\end{table}

\begin{figure}[ht]
    % \centering
    {
        \includegraphics[width=\linewidth]{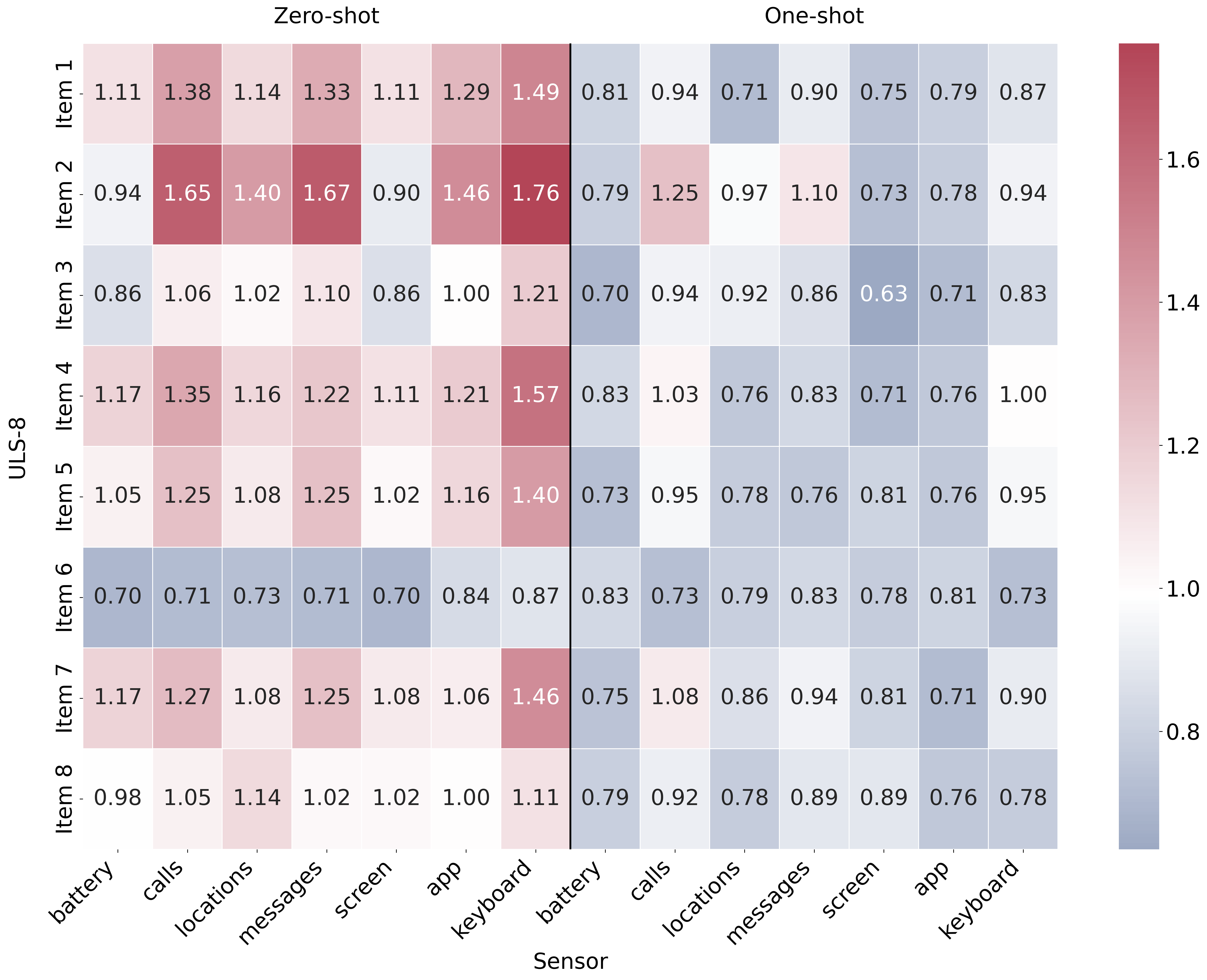}
    }
    \vspace{-5mm}
    \caption{Mean Absolute Error (MAE) of zero-shot and one-shot inferences for predicting ULS-8 item scores across different smartphone sensors.}
    \label{heatmap}
\end{figure}

As shown in Figure \ref{heatmap}, one-shot inference significantly reduced the prediction errors for most sub-questions of ULS-8, but the degree of improvement varied substantially. For keyboard sensors, the MAE for Item 2 (``There is no one I can turn to'') reduced most significantly ($\Delta \text{MAE}_{\text{keyboard},\text{Item2}}$ = -46.8\%), while Item 6 (``I can find companionship when I want it'') showed the smallest improvement ($\Delta \text{MAE}_{\text{app},\text{Item6}}$ = -3.8\%). In contrast, the keyboard sensor showed consistent improvement, with the MAE for all individual items reduced by over 16\%, and particularly notable reductions for Item 1 and Item 2 ($\Delta \text{MAE}_{\text{keyboard},\text{Item1}}$ = -41.5\%, $\Delta \text{MAE}_{\text{keyboard},\text{Item2}}$ =  -46.8\%). Notably, Item 6 exhibited error increases on battery usage ($\Delta \text{MAE}_{\text{battery},\text{Item6}}$ = +18.2\%) and locations ($\Delta \text{MAE}_{\text{locations},\text{Item6}}$ = +8.7\%) sensors, reflecting the limitations of prior knowledge about these two types of smartphone sensors in enhancing individuals' ability to perceive their capacity for obtaining companionship on demand and their predictive ability for convenience. 

Overall, as Table \ref{tab:sensor_performance} illustrates, one-shot mode predicting ULS-8 total score reduced the MAE of all sensors by 32.8\% averagely comparing to zero-shot results, with the largest contributions from keyboard activity sensors ($\Delta \text{MAE}_{\text{keyboard}}$ = -42.3\%) and application usage sensors ($\Delta \text{MAE}_{\text{app}}$ = -39.9\%), while message interaction ($\Delta \text{MAE}_{\text{messages}}$ = -33.2\%) and screen usage ($\Delta \text{MAE}_{\text{screen}}$ = -28.1\%) sensors showed relatively smaller improvements. This pattern indicates that items related to social help-seeking (Item 2 - There is no one I can turn to) and extroverted cognition (Item 3 - I am an outgoing person) benefit the most from contextual learning, whereas predicting companionship availability (Item 6 - I can find companionship when I want it) requires more sophisticated sensor fusion strategies.

In terms of MBE, the one-shot setting also led to consistently lower bias across all sensor modalities, suggesting more balanced estimations rather than systematic over- or under-prediction.

When predicting ULS-8 total score, the most notable MBE reductions were observed in application usage ($\Delta \text{MBE}_{\text{app}}$ = -53.4\%) and screen activity ($\Delta \text{MBE}_{\text{screen}}$ = -53.4\%), implying that contextual examples helped correct directional prediction errors in these domains (Table \ref{tab:sensor_performance}). Other modalities such as battery ($\Delta \text{MBE}_{\text{battery}}$ = -46.6\%) and keyboard usage ($\Delta \text{MBE}_{\text{keyboard}}$ = -53.3\%) also benefited significantly. This reduction in bias aligns with the observed gains in MAE and further underscores the advantage of contextual prompting in improving both the accuracy and reliability of LLM-based inferences in predicting levels of loneliness from smartphone sensor data.

Additionally, in both zero-shot and one-shot settings, the mean bias error (MBE) between the ULS-8 total scores predicted by each sensor and the ground truth values was positive, indicating the tendency of LLMs overestimating users' loneliness scores when encountering unseen or few examples.

Analyzing the reasoning results produced by LLMs, we found that LLM reasoning aligned behavioral patterns with key social dimensions: social engagement (screen time and social apps), communication stability and support (calls and messages), and digital interaction levels (locations and keyboard activity). We also found the LLM’s reliance on conservative scoring when observing missing data. The model also exhibited epistemic caution in prediction. We also observed phrases like ``not conclusive evidence'' and references to ``limited interaction'' suggest that incomplete or sparse sensor data prompted conservative scoring, avoiding extreme ``never'' or ``often'' classifications.

\subsection{Generalized Random Forest Models for Predicting Loneliness Scores}

\begin{figure}[ht]
  \centering
  \subfigure{
    \includegraphics[trim=0pt 9.5pt 0pt 0pt, clip, width=0.22\textwidth]{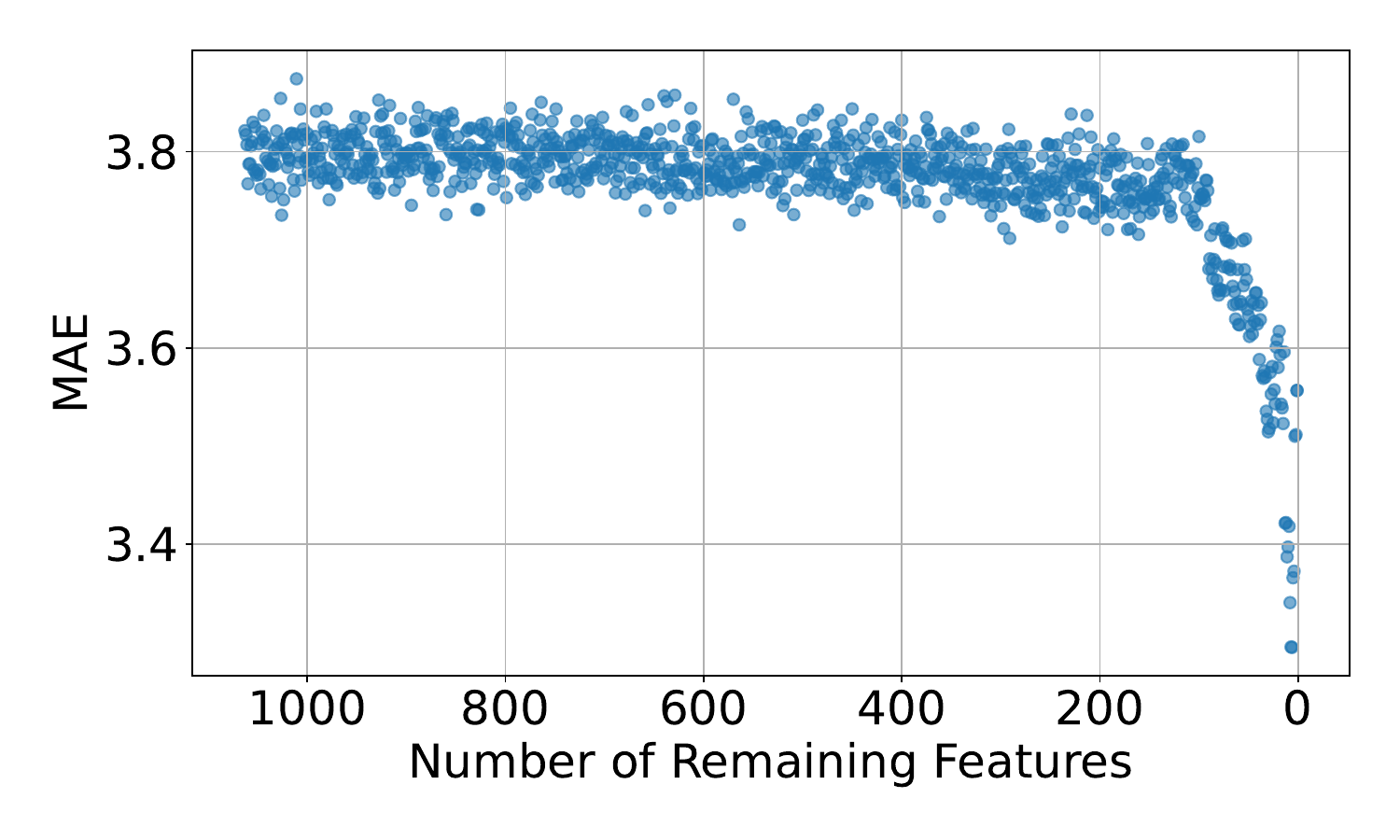}
  }
  \subfigure{
    \includegraphics[trim=0pt 9.5pt 0pt 0pt, clip, width=0.22\textwidth]{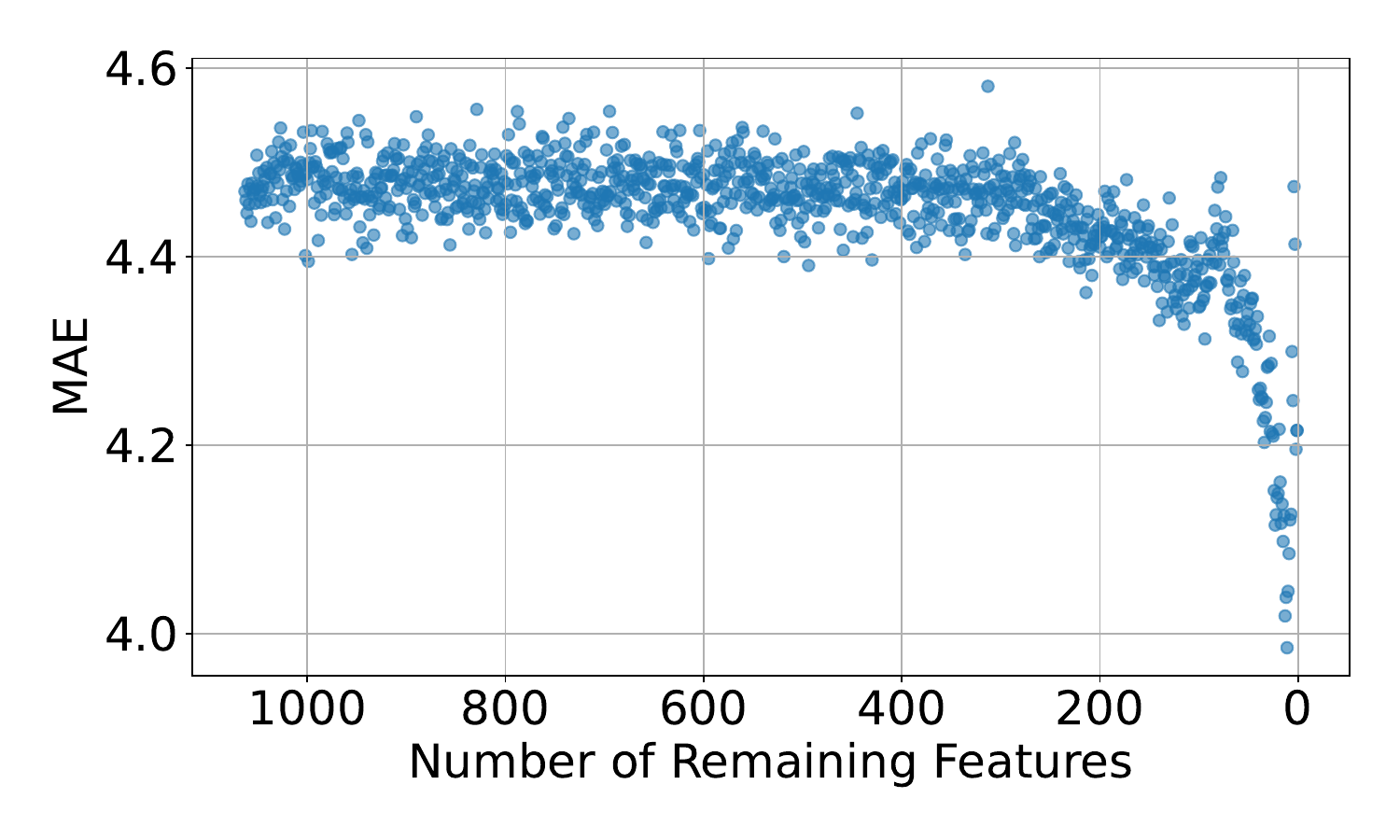}
  }
  \vspace{-5mm}
  \caption{Dot plot distribution of remaining feature count from smartphone sensors vs. Cross-Validation MAE in ULS-8 Loneliness prediction models at midterm (left) and end-of-semester (right).}
  \label{dot_plot_completion}
\end{figure}

\begin{figure}[ht]
  \centering
  \subfigure{
    \includegraphics[trim=0pt 9pt 0pt 0pt, clip, width=0.22\textwidth]{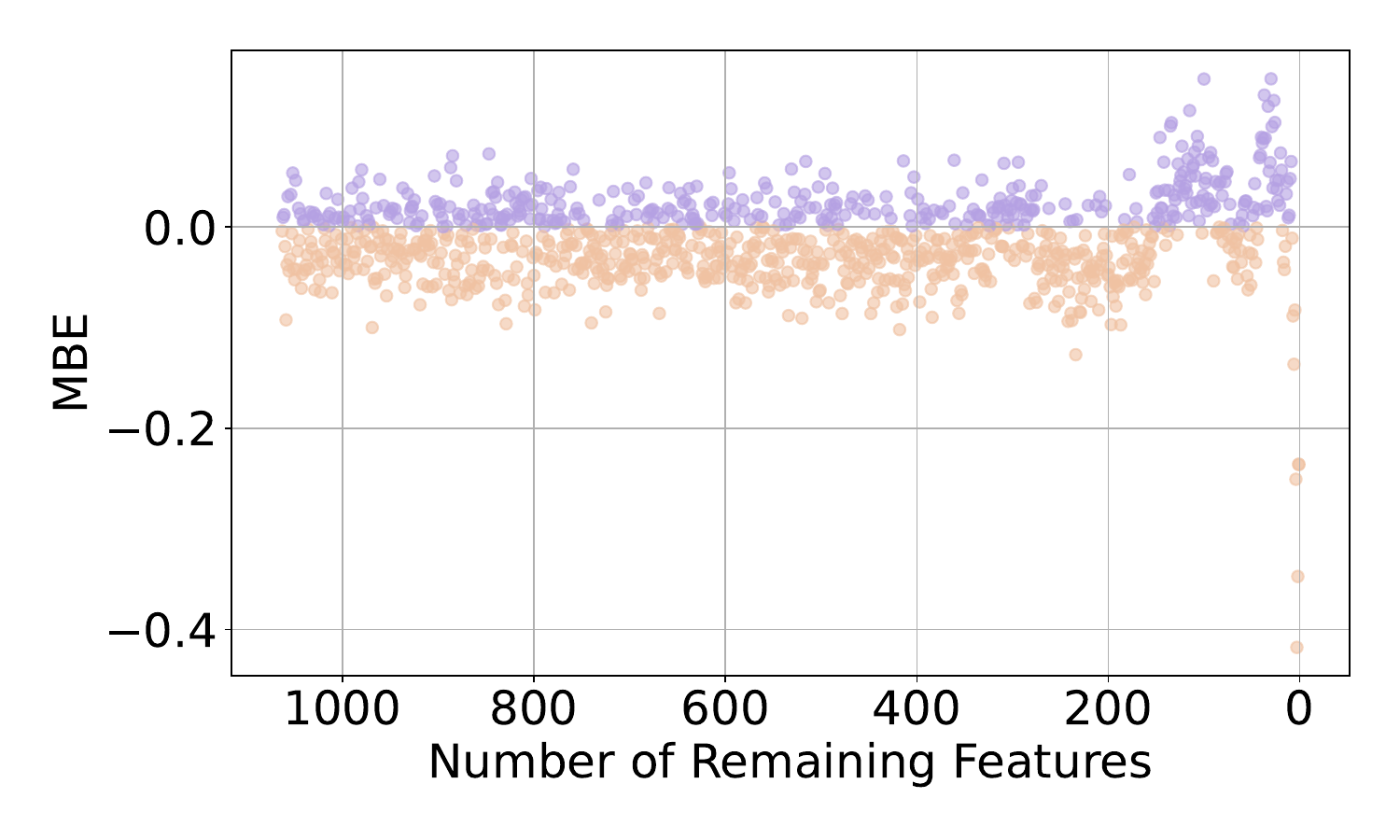}
  }
  \subfigure{
    \includegraphics[trim=0pt 9pt 0pt 0pt, clip, width=0.22\textwidth]{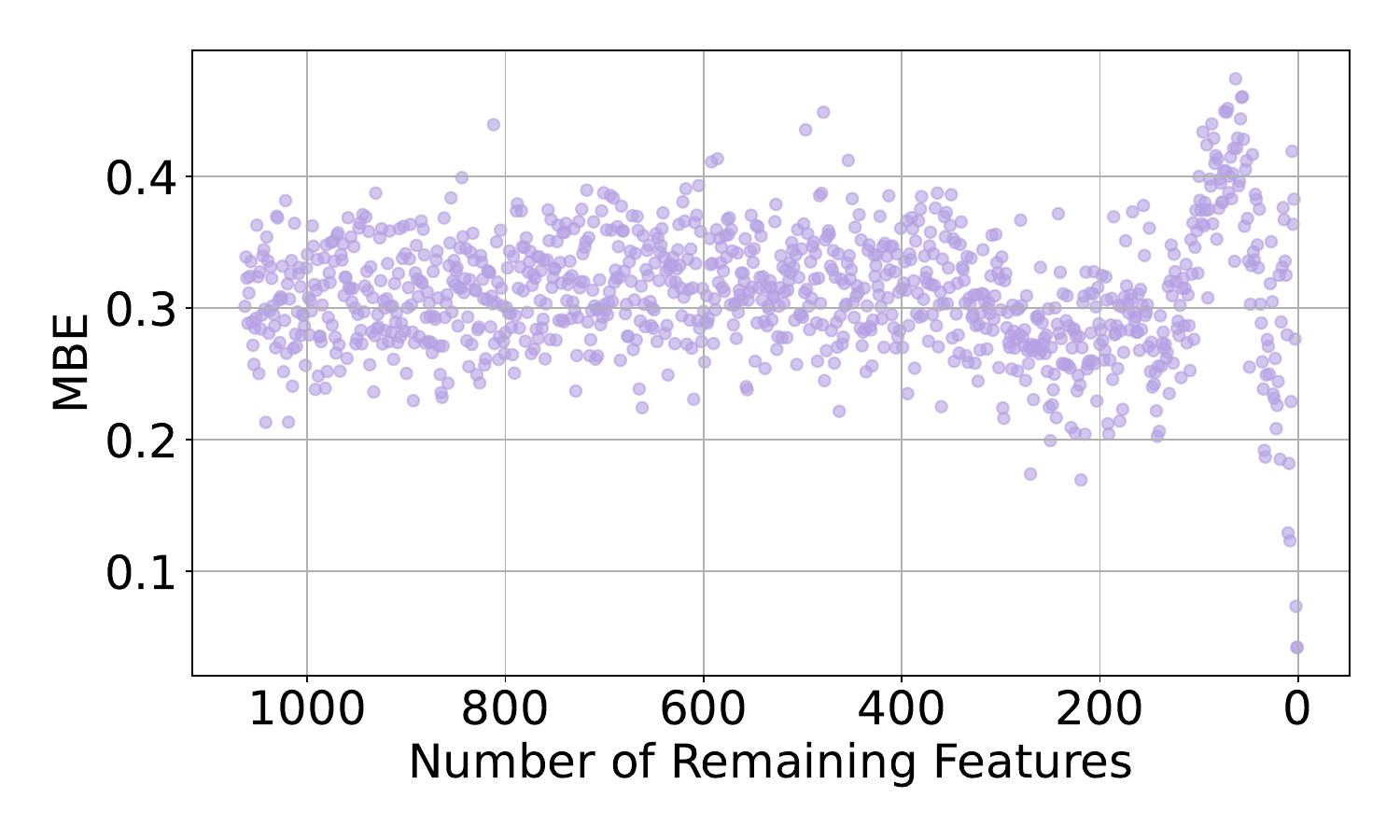}
  }
  \vspace{-5mm}
  \caption{Dot plot distribution of remaining feature count from smartphone sensors vs. Cross-Validation MBE in ULS-8 Loneliness prediction models at midterm (left) and end-of-semester (right).}
  \label{dot_plot_completion_MBE}
\end{figure}

The dynamic relationship between the number of features and prediction error reveals a significant threshold effect. As illustrated in Figures \ref{dot_plot_completion}, the MAE in the midterm and end-of-semester stages reaches minimum values of 3.29 and 3.98 when the remaining feature counts are 6 and 11 respectively. These retained features represent the set of best-performing smartphone sensor features predicting university students' reported loneliness. The similarity in the effective feature set size implies that the core behavioral signals associated with loneliness may remain consistent at midterm and the end of semester, underlying behavioral regularities of university students.

We observed that at both midterm and the end of semester, the MBE evaluated in generalized models remained relatively stable around zero as progressively reducing sensor features (Figure \ref{dot_plot_completion_MBE}). This indicates models trained with the best-performing smartphone sensing features yield more stable predictions and higher-dimensional input features may lead to overestimation.

% \vspace{-3mm}
\subsubsection{\textbf{Predictive Features}} 

Despite the large number of features initially considered, the optimal subsets identified by the random forest models at both the midterm and end of semester (Tables \ref{midterm_best_features} and \ref{completion_best_features}) reveal consistent sensor-level patterns. Notably, screen emerged as the core modality in both midterm and the end-of-semester models, highlighting the sustained predictive value of smartphone usage duration for loneliness. 

In the midterm stage, the optimal features for predicting loneliness selected are highly concentrated in the Screen sensor (especially indicators related to screen unlock duration), highlighting the core role of mobile phone usage time in midterm predictions. By the end of semester, features from the Screen sensor remain dominant but become more diversified (e.g. First Unlock Time After Midnight on Day 6 \& 14 and Standard Deviation of Unlock Duration on Day 12). Additionally, the optimal feature set incorporates key features from the Locations sensor (e.g. Moving to Static Ratio on Day 3 \& 8) and Battery sensor (e.g. Total Discharge Duration on Day 6). This indicates that as the semester progresses, the smartphone sensor features most predictive of loneliness have evolved from a single focus on mobile phone usage intensity in the midterm to a complex combination at the end of semester, integrating regularity of activity patterns, degree of social isolation, and multi-dimensional mobile phone usage habits. 

Notably, real-time interaction sensors such as applications, calls, and messages were not in the core feature set at the end of semester. This suggests that social activity data has limited predictive validity for loneliness among university students at the end of semester.

\subsubsection{\textbf{Temporal Distribution Patterns}}

The temporal distribution of sensor features revealed distinct aggregation patterns at midterm and the end of semester, which closely associated with the behavioral patterns of university students. The optimal features were relatively evenly distributed across the 14-day observation period, without obvious concentration on specific days. Such evenness suggests that behavioral indicators related to loneliness may persist over time rather than peak at specific moments.

Additionally, this study found no significant patterns indicating whether these features tended to cluster between weekdays and weekends or between the first and second weeks. Individual differences in academic schedules and lifestyles might have introduced noise into the temporal distribution, especially given the diversity of majors among our student participants. There is also a potential limitation of relatively small sample size, which may have obscured subtle temporal trends. Future studies could align feature analysis more closely with personalized academic calendars or event-based timelines and incorporate larger and more diverse samples to further validate the temporal distribution of significant features.

\begin{table}[ht]
\centering
\scriptsize
\caption{Smartphone sensing features most predictive of ULS-8 total scores at midterm using a Random Forest Model (N = 88, MAE = 3.29), sorted alphabetically.}
\label{midterm_best_features}
\begin{tabular}{lll}
\hline
\textbf{Number} & \textbf{Sensor} & \textbf{Feature} \\
\hline
1 & Locations & Average Speed (Day 8) \\ \midrule
2 & Screen & Average Unlock Duration (Day 13) \\
3 & Screen & Minimum Unlock Duration (Day 4) \\
4 & Screen & Minimum Unlock Duration (Day 13) \\
5 & Screen & Total Unlock Duration (Day 3) \\
6 & Screen & Total Unlock Duration (Day 8) \\
\hline
\end{tabular}
\end{table}

\begin{table}[ht]
\centering
\scriptsize
\caption{Smartphone sensing features most predictive of ULS-8 total scores at the end of semester using a Random Forest Model (N = 88, MAE = 3.98), sorted alphabetically}
\label{completion_best_features}
\begin{tabular}{lll}
\hline
\textbf{Number} & \textbf{Sensor} & \textbf{Feature} \\
\hline
1 & Battery & Total Discharge Duration (Day 6) \\ \midrule
2 & Keyboard & Average Inter-key Delay (Day 3) \\ \midrule
3 & Locations & Moving to Static Ratio (Day 3) \\
4 & Locations & Moving to Static Ratio (Day 8) \\
5 & Locations & Standard Deviation of Stay Length at Clusters (Day 7) \\ \midrule
6 & Screen & First Unlock Time After Midnight (Day 6) \\
7 & Screen & First Unlock Time After Midnight (Day 14) \\
8 & Screen & Standard Deviation of Unlock Duration (Day 12) \\
9 & Screen & Total Unlock Duration (Day 5) \\
10 & Screen & Unlock Episode Count (Day 5) \\
11 & Screen & Unlock Episode Count (Day 10) \\
\hline
\end{tabular}
\end{table}

\subsection{Insights from Predictive Models for University Students' Reported Loneliness Levels}

\subsubsection{\textbf{Discussion}} At midterm and the end of semester, the six core features shown in Table \ref{midterm_best_features} and the eleven core features shown in Table \ref{completion_best_features} can accurately predict loneliness among university students because they collectively reveal systematic changes in individuals' daily behaviors during these two 14-day periods of the semester. The selected features collectively capture the dynamic generative patterns of loneliness across the weeks. These features corroborate earlier studies \cite{saeb2016mobile, wang2018tracking, enez2016smartphone, wang2014studentlife} which found the following smartphone sensing features as key indicators of loneliness: increased screen usage time, reduced location transitions, increased late-night phone usage, and prolonged stay in a single location. Also, previous work indicated that elevated smartphone usage may reflect attempts at self-medication by individuals trying to cope with feeling lonely when troubled, although not necessarily reducing loneliness \cite{enez2016smartphone, orben2019social}. Therefore, the passively collected smartphone sensing data offer a reliable source for identifying loneliness-related behaviors.

Our findings indicate that minimal-context learning significantly enhances LLMs' mental health prediction capabilities. Specifically, when predicting ULS-8 scores from smartphone sensor data, the one-shot prediction results outperformed those of the zero-shot mode. This aligns with the core conclusion of a previous study \citep{xu2024mental, zhang2024leveraging} which demonstrated that few-shot prompting examples substantially improve model accuracy across diverse mental health tasks.

In our study, the emotional, social, and global dimensions of loneliness showed notable differences in prediction accuracy when using LLMs. Specifically, items related to emotional loneliness (e.g., Item 4 - I feel excluded) showed lower prediction accuracy (39.68\%). This aligns with Ozdemir and Tan's \citep{ozdemir2024meta} insight that emotional loneliness typically involves internal emotions and subjective experiences, which smartphone sensor data struggle to fully capture due to its limitation in recording nuanced affective changes. By contrast, social loneliness-related items (e.g., Item 2 - I can find companionship) achieved higher accuracy (41.27\%). This is consistent with the characteristics of social loneliness documented in a previous study \citep{wu2021improving}, which suggests that the results may because social loneliness is more tied to external behaviors such as social interactions, daily activities, and behavioral patterns, which are more readily captured by smartphone sensors, thereby enhancing prediction accuracy.

Moreover, a study \citep{enez2016smartphone} exploring the relationship between smartphone use and social psychology has found that usage data from smartphones play a crucial role in loneliness prediction, consistent with the sensor data employed in our research. For instance, features like screen unlock duration and unlock episode count emerged as key predictors in our optimal feature set for the generalized models. The screen sensor to which they belong reflects the frequency of mobile phone use and has become the sensor with the smallest mean bias error in LLM's prediction of loneliness among university students, particularly in items like Item 3 and Item 6. Also, analogous behavioral metrics such as location transitions and stay duration demonstrated significant predictive value, elaborating previous findings \citep{jafarlou2024objective}.

\vspace{-2mm}
\subsubsection{\textbf{Key Highlights}}This study demonstrates that machine learning models trained on population-level smartphone sensing data can accurately predict ULS-8 loneliness scores in university students by capturing shared behavioral patterns, while personalized models offer practical adaptability in data-sparse situations by inferring loneliness without retraining.

Using personalized LLM-based inference on smartphone sensing data, application usage achieved the highest one-shot prediction performance for ULS-8 loneliness. Smartphone sensors keyboard and applications showed the greatest improvement from zero-shot to one-shot learning.

Among the best-performing machine learning models for predicting ULS-8 loneliness scores, screen sensor features were the most predictive at both midterm and end-of-semester. The best predictive features of loneliness were dominated by the features of the screen and locations sensor.

\vspace{-2mm}
\subsubsection{\textbf{Limitation}}

This study's methodology has several limitations. While the sample size was adequate for the scope of the current scope, it may be insufficient to generalize findings across broader and more diverse populations. Also, the analysis of smartphone sensor features relied on traditional numeric statistics rather than contextual interpretations. Future work could address these limitations by engaging larger and more diverse cohorts and exploring more naunced approaches to sensor data analysis.

In terms of privacy and interpretability, personalized models powered by LLMs can offer user-specific explanations and preserve privacy, particularly when implemented on-device. In contrast, while machine learning models trained on broader datasets often achieve higher accuracy, they are less interpretable at the individual level. There is an inherent trade-off between personalization and performance, which should be carefully considered as these approaches continue to evolve.

\section{Future Work and Conclusions}
In conclusion, our study leveraged generalized machine learning and personalized LLM inference to predict university students' loneliness using smartphone sensing data. Our findings show that both generalized machine learning models and LLM-based personalized approaches offer distinct predictive value: the former provides interpretable population-level insights, identifying cross-sensor features like screen using duration and location mobility dynamically correlated with loneliness scores; the latter demonstrates superior individual-level prediction via one-shot learning, significantly improving accuracy for keyboard input and app usage data.

Future research could expand to more diverse student populations and long-term behavioral data, as well as integrating contextual information (e.g., academic schedules, smartphone text content) to enhance model robustness and personalization.

\bibliographystyle{ACM-Reference-Format}
\balance
\bibliography{Reference.bib}


\begin{thebibliography}{19}

%%% the Web address.

\ifx \showCODEN    \undefined \def \showCODEN     #1{\unskip}     \fi
\ifx \showISBNx    \undefined \def \showISBNx     #1{\unskip}     \fi
\ifx \showISBNxiii \undefined \def \showISBNxiii  #1{\unskip}     \fi
\ifx \showISSN     \undefined \def \showISSN      #1{\unskip}     \fi
\ifx \showLCCN     \undefined \def \showLCCN      #1{\unskip}     \fi
\ifx \shownote     \undefined \def \shownote      #1{#1}          \fi
\ifx \showarticletitle \undefined \def \showarticletitle #1{#1}   \fi
\ifx \showURL      \undefined \def \showURL       {\relax}        \fi
% invisible to TeX
\providecommand\bibfield[2]{#2}
\providecommand\bibinfo[2]{#2}
\providecommand\natexlab[1]{#1}
\providecommand\showeprint[2][]{arXiv:#2}

\bibitem[Barreto et~al\mbox{.}(2021)]%
        {barreto2021loneliness}
\bibfield{author}{\bibinfo{person}{Manuela Barreto}, \bibinfo{person}{Christina Victor}, \bibinfo{person}{Claudia Hammond}, \bibinfo{person}{Alice Eccles}, \bibinfo{person}{Matt~T Richins}, {and} \bibinfo{person}{Pamela Qualter}.} \bibinfo{year}{2021}\natexlab{}.
\newblock \showarticletitle{Loneliness around the world: Age, gender, and cultural differences in loneliness}.
\newblock \bibinfo{journal}{\emph{Personality and individual differences}}  \bibinfo{volume}{169} (\bibinfo{year}{2021}), \bibinfo{pages}{110066}.
\newblock


\bibitem[Doryab et~al\mbox{.}(2019)]%
        {doryab2019identifying}
\bibfield{author}{\bibinfo{person}{Afsaneh Doryab}, \bibinfo{person}{Daniella~K Villalba}, \bibinfo{person}{Prerna Chikersal}, \bibinfo{person}{Janine~M Dutcher}, \bibinfo{person}{Michael Tumminia}, \bibinfo{person}{Xinwen Liu}, \bibinfo{person}{Sheldon Cohen}, \bibinfo{person}{Kasey Creswell}, \bibinfo{person}{Jennifer Mankoff}, \bibinfo{person}{John~D Creswell}, {et~al\mbox{.}}} \bibinfo{year}{2019}\natexlab{}.
\newblock \showarticletitle{Identifying behavioral phenotypes of loneliness and social isolation with passive sensing: statistical analysis, data mining and machine learning of smartphone and fitbit data}.
\newblock \bibinfo{journal}{\emph{JMIR mHealth and uHealth}} \bibinfo{volume}{7}, \bibinfo{number}{7} (\bibinfo{year}{2019}), \bibinfo{pages}{e13209}.
\newblock


\bibitem[Enez~Darcin et~al\mbox{.}(2016)]%
        {enez2016smartphone}
\bibfield{author}{\bibinfo{person}{Asli Enez~Darcin}, \bibinfo{person}{Samet Kose}, \bibinfo{person}{Cemal~Onur Noyan}, \bibinfo{person}{Serdar Nurmedov}, \bibinfo{person}{Onat Y{\i}lmaz}, {and} \bibinfo{person}{Nesrin Dilbaz}.} \bibinfo{year}{2016}\natexlab{}.
\newblock \showarticletitle{Smartphone addiction and its relationship with social anxiety and loneliness}.
\newblock \bibinfo{journal}{\emph{Behaviour \& Information Technology}} \bibinfo{volume}{35}, \bibinfo{number}{7} (\bibinfo{year}{2016}), \bibinfo{pages}{520--525}.
\newblock


\bibitem[Harari et~al\mbox{.}(2017)]%
        {harari2017smartphone}
\bibfield{author}{\bibinfo{person}{Gabriella~M Harari}, \bibinfo{person}{Sandrine~R M{\"u}ller}, \bibinfo{person}{Min~SH Aung}, {and} \bibinfo{person}{Peter~J Rentfrow}.} \bibinfo{year}{2017}\natexlab{}.
\newblock \showarticletitle{Smartphone sensing methods for studying behavior in everyday life}.
\newblock \bibinfo{journal}{\emph{Current opinion in behavioral sciences}}  \bibinfo{volume}{18} (\bibinfo{year}{2017}), \bibinfo{pages}{83--90}.
\newblock


\bibitem[Hays and DiMatteo(1987)]%
        {hays1987short}
\bibfield{author}{\bibinfo{person}{Ron~D Hays} {and} \bibinfo{person}{M~Robin DiMatteo}.} \bibinfo{year}{1987}\natexlab{}.
\newblock \showarticletitle{A short-form measure of loneliness}.
\newblock \bibinfo{journal}{\emph{Journal of personality assessment}} \bibinfo{volume}{51}, \bibinfo{number}{1} (\bibinfo{year}{1987}), \bibinfo{pages}{69--81}.
\newblock


\bibitem[Jafarlou et~al\mbox{.}(2024)]%
        {jafarlou2024objective}
\bibfield{author}{\bibinfo{person}{Salar Jafarlou}, \bibinfo{person}{Iman Azimi}, \bibinfo{person}{Jocelyn Lai}, \bibinfo{person}{Yuning Wang}, \bibinfo{person}{Sina Labbaf}, \bibinfo{person}{Brenda Nguyen}, \bibinfo{person}{Hana Qureshi}, \bibinfo{person}{Christopher Marcotullio}, \bibinfo{person}{Jessica~L Borelli}, \bibinfo{person}{Nikil~D Dutt}, {et~al\mbox{.}}} \bibinfo{year}{2024}\natexlab{}.
\newblock \showarticletitle{Objective monitoring of loneliness levels using smart devices: A multi-device approach for mental health applications}.
\newblock \bibinfo{journal}{\emph{Plos one}} \bibinfo{volume}{19}, \bibinfo{number}{6} (\bibinfo{year}{2024}), \bibinfo{pages}{e0298949}.
\newblock


\bibitem[Onnela and Rauch(2016)]%
        {onnela2016harnessing}
\bibfield{author}{\bibinfo{person}{Jukka-Pekka Onnela} {and} \bibinfo{person}{Scott~L Rauch}.} \bibinfo{year}{2016}\natexlab{}.
\newblock \showarticletitle{Harnessing smartphone-based digital phenotyping to enhance behavioral and mental health}.
\newblock \bibinfo{journal}{\emph{Neuropsychopharmacology}} \bibinfo{volume}{41}, \bibinfo{number}{7} (\bibinfo{year}{2016}), \bibinfo{pages}{1691--1696}.
\newblock


\bibitem[Orben and Przybylski(2019)]%
        {orben2019social}
\bibfield{author}{\bibinfo{person}{Amy Orben} {and} \bibinfo{person}{Andrew~K Przybylski}.} \bibinfo{year}{2019}\natexlab{}.
\newblock \showarticletitle{The association between adolescent well-being and digital technology use}.
\newblock \bibinfo{journal}{\emph{Nature Human Behaviour}} \bibinfo{volume}{3}, \bibinfo{number}{2} (\bibinfo{year}{2019}), \bibinfo{pages}{173--182}.
\newblock


\bibitem[Ozdemir and Tan(2024)]%
        {ozdemir2024meta}
\bibfield{author}{\bibinfo{person}{Vildan Ozdemir} {and} \bibinfo{person}{Seref Tan}.} \bibinfo{year}{2024}\natexlab{}.
\newblock \showarticletitle{Meta-analytic factor analysis of the UCLA loneliness scale}.
\newblock \bibinfo{journal}{\emph{Current Psychology}} \bibinfo{volume}{43}, \bibinfo{number}{20} (\bibinfo{year}{2024}), \bibinfo{pages}{18307--18318}.
\newblock


\bibitem[Qirtas et~al\mbox{.}(2023)]%
        {qirtas2023personalizing}
\bibfield{author}{\bibinfo{person}{Malik~Muhammad Qirtas}, \bibinfo{person}{Eleanor~Bantry White}, \bibinfo{person}{Evi Zafeiridi}, {and} \bibinfo{person}{Dirk Pesch}.} \bibinfo{year}{2023}\natexlab{}.
\newblock \showarticletitle{Personalizing loneliness detection through behavioral grouping of passive sensing data from college students}.
\newblock \bibinfo{journal}{\emph{IEEE Access}}  \bibinfo{volume}{11} (\bibinfo{year}{2023}), \bibinfo{pages}{88841--88851}.
\newblock


\bibitem[Saeb et~al\mbox{.}(2015)]%
        {saeb2016mobile}
\bibfield{author}{\bibinfo{person}{Sohrab Saeb}, \bibinfo{person}{Mo Zhang}, \bibinfo{person}{Christopher~J Karr}, \bibinfo{person}{Stephen~M Schueller}, \bibinfo{person}{Marya~E Corden}, \bibinfo{person}{Konrad~P Kording}, {and} \bibinfo{person}{David~C Mohr}.} \bibinfo{year}{2015}\natexlab{}.
\newblock \showarticletitle{Mobile phone sensor correlates of depressive symptom severity in daily-life behavior: An exploratory study}.
\newblock \bibinfo{journal}{\emph{Journal of medical Internet research}} \bibinfo{volume}{17}, \bibinfo{number}{7} (\bibinfo{year}{2015}), \bibinfo{pages}{e175}.
\newblock


\bibitem[van Berkel et~al\mbox{.}(2023)]%
        {van2023aware}
\bibfield{author}{\bibinfo{person}{Niels van Berkel}, \bibinfo{person}{Simon D’Alfonso}, \bibinfo{person}{Rio Kurnia~Susanto}, \bibinfo{person}{Denzil Ferreira}, {and} \bibinfo{person}{Vassilis Kostakos}.} \bibinfo{year}{2023}\natexlab{}.
\newblock \showarticletitle{AWARE-Light: A smartphone tool for experience sampling and digital phenotyping}.
\newblock \bibinfo{journal}{\emph{Personal and Ubiquitous Computing}} \bibinfo{volume}{27}, \bibinfo{number}{2} (\bibinfo{year}{2023}), \bibinfo{pages}{435--445}.
\newblock


\bibitem[Vega et~al\mbox{.}(2021)]%
        {rapids}
\bibfield{author}{\bibinfo{person}{Julio Vega}, \bibinfo{person}{Meng Li}, \bibinfo{person}{Kwesi Aguillera}, \bibinfo{person}{Nikunj Goel}, \bibinfo{person}{Echhit Joshi}, \bibinfo{person}{Kirtiraj Khandekar}, \bibinfo{person}{Krina~C. Durica}, \bibinfo{person}{Abhineeth~R. Kunta}, {and} \bibinfo{person}{Carissa~A. Low}.} \bibinfo{year}{2021}\natexlab{}.
\newblock \showarticletitle{Reproducible Analysis Pipeline for Data Streams: Open-Source Software to Process Data Collected With Mobile Devices}.
\newblock \bibinfo{journal}{\emph{Frontiers in Digital Health}}  \bibinfo{volume}{3} (\bibinfo{year}{2021}).
\newblock
\showISSN{2673-253X}
\href{https://doi.org/10.3389/fdgth.2021.769823}{doi:\nolinkurl{10.3389/fdgth.2021.769823}}


\bibitem[Wang et~al\mbox{.}(2014)]%
        {wang2014studentlife}
\bibfield{author}{\bibinfo{person}{Rui Wang}, \bibinfo{person}{Fanglin Chen}, \bibinfo{person}{Zhenyu Chen}, \bibinfo{person}{Ting Li}, \bibinfo{person}{Gabriella Harari}, \bibinfo{person}{Sarah Tignor}, \bibinfo{person}{Xia Zhou}, \bibinfo{person}{Dror Ben-Zeev}, {and} \bibinfo{person}{Andrew~T Campbell}.} \bibinfo{year}{2014}\natexlab{}.
\newblock \showarticletitle{StudentLife: assessing mental health, academic performance and behavioral trends of college students using smartphones}. In \bibinfo{booktitle}{\emph{Proceedings of the 2014 ACM International Joint Conference on Pervasive and Ubiquitous Computing}}. ACM, \bibinfo{pages}{3--14}.
\newblock


\bibitem[Wang et~al\mbox{.}(2018)]%
        {wang2018tracking}
\bibfield{author}{\bibinfo{person}{Rui Wang}, \bibinfo{person}{Wenchao Wang}, \bibinfo{person}{Aileen daSilva}, \bibinfo{person}{Jeremy~F Huckins}, \bibinfo{person}{William~M Kelley}, \bibinfo{person}{Todd~F Heatherton}, {and} \bibinfo{person}{Andrew~T Campbell}.} \bibinfo{year}{2018}\natexlab{}.
\newblock \showarticletitle{Tracking depression dynamics in college students using mobile phone and wearable sensing}.
\newblock \bibinfo{journal}{\emph{Proceedings of the ACM on Interactive, Mobile, Wearable and Ubiquitous Technologies}} \bibinfo{volume}{2}, \bibinfo{number}{1} (\bibinfo{year}{2018}), \bibinfo{pages}{1--26}.
\newblock


\bibitem[Wu et~al\mbox{.}(2021)]%
        {wu2021improving}
\bibfield{author}{\bibinfo{person}{Congyu Wu}, \bibinfo{person}{Amanda~N Barczyk}, \bibinfo{person}{R~Cameron Craddock}, \bibinfo{person}{Gabriella~M Harari}, \bibinfo{person}{Edison Thomaz}, \bibinfo{person}{Jason~D Shumake}, \bibinfo{person}{Christopher~G Beevers}, \bibinfo{person}{Samuel~D Gosling}, {and} \bibinfo{person}{David~M Schnyer}.} \bibinfo{year}{2021}\natexlab{}.
\newblock \showarticletitle{Improving prediction of real-time loneliness and companionship type using geosocial features of personal smartphone data}.
\newblock \bibinfo{journal}{\emph{Smart Health}}  \bibinfo{volume}{20} (\bibinfo{year}{2021}), \bibinfo{pages}{100180}.
\newblock


\bibitem[Xu et~al\mbox{.}(2024)]%
        {xu2024mental}
\bibfield{author}{\bibinfo{person}{Xuhai Xu}, \bibinfo{person}{Bingsheng Yao}, \bibinfo{person}{Yuanzhe Dong}, \bibinfo{person}{Saadia Gabriel}, \bibinfo{person}{Hong Yu}, \bibinfo{person}{James Hendler}, \bibinfo{person}{Marzyeh Ghassemi}, \bibinfo{person}{Anind~K Dey}, {and} \bibinfo{person}{Dakuo Wang}.} \bibinfo{year}{2024}\natexlab{}.
\newblock \showarticletitle{Mental-llm: Leveraging large language models for mental health prediction via online text data}.
\newblock \bibinfo{journal}{\emph{Proceedings of the ACM on Interactive, Mobile, Wearable and Ubiquitous Technologies}} \bibinfo{volume}{8}, \bibinfo{number}{1} (\bibinfo{year}{2024}), \bibinfo{pages}{1--32}.
\newblock


\bibitem[Yang et~al\mbox{.}(2023)]%
        {yang2023loneliness}
\bibfield{author}{\bibinfo{person}{Zhongqi Yang}, \bibinfo{person}{Iman Azimi}, \bibinfo{person}{Salar Jafarlou}, \bibinfo{person}{Sina Labbaf}, \bibinfo{person}{Jessica Borelli}, \bibinfo{person}{Nikil Dutt}, {and} \bibinfo{person}{Amir~M Rahmani}.} \bibinfo{year}{2023}\natexlab{}.
\newblock \showarticletitle{Loneliness forecasting using multi-modal wearable and mobile sensing in everyday settings}. In \bibinfo{booktitle}{\emph{2023 IEEE 19th International Conference on Body Sensor Networks (BSN)}}. IEEE, \bibinfo{pages}{1--4}.
\newblock


\bibitem[Zhang et~al\mbox{.}(2024)]%
        {zhang2024leveraging}
\bibfield{author}{\bibinfo{person}{Tianyi Zhang}, \bibinfo{person}{Songyan Teng}, \bibinfo{person}{Hong Jia}, {and} \bibinfo{person}{Simon D'Alfonso}.} \bibinfo{year}{2024}\natexlab{}.
\newblock \showarticletitle{Leveraging LLMs to predict affective states via smartphone sensor features}. In \bibinfo{booktitle}{\emph{Companion of the 2024 on ACM International Joint Conference on Pervasive and Ubiquitous Computing}}. \bibinfo{pages}{709--716}.
\newblock


\end{thebibliography}

% \newpage
% \appendix 
% \input{Chapter/Appendix}

\end{document}